\pdfoutput=1
\documentclass[11pt,twoside,a4paper,cmspaper,final,collab]{cms-tdr}
\def\svnVersion{cc78786-D}\def\svnDate{2021/09/21}\def\cmsCernNoTag{CERN-EP-2020-196}\def\cmsCernDate{\today}\def\cmsMessage{"Published in Physical Review Letters as \href{http://dx.doi.org/10.1103/PhysRevLett.127.122001}{\doi{10.1103/PhysRevLett.127.122001}.}"}
\begin{document}\cmsNoteHeader{HIN-19-014}

\newcommand {\PbPb}  {\ensuremath{\text{Pb-Pb}}\xspace}
\newcommand {\TwoGamma}  {\ensuremath{\PGg\PGg}\xspace}
\newcommand {\GGMuMu}  {\ensuremath{\PGg\PGg \to \PGmp\PGmm}\xspace}
\newcommand {\MeanPt}  {\ensuremath{\langle \pt \rangle}\xspace}
\newcommand {\MeanAlphaCore}  {\ensuremath{\langle \alpha^\text{core} \rangle}\xspace}
\newcommand {\MeanMass}  {\ensuremath{\langle m_{\PGm\PGm} \rangle}\xspace}
\newcommand {\trigEff}  {\ensuremath{\varepsilon_{\text{trig}}}\xspace}
\newcommand {\recoEff}  {\ensuremath{\varepsilon_{\text{reco}}}\xspace}

\ifthenelse{\boolean{cms@external}}{\providecommand{\suppMaterial}{available in the Supplemental Material~\cite{suppMat}}}{\providecommand{\suppMaterial}{see Appendix~\ref{app:suppMat}\nocite{suppMat, Zyla:2020zbs}}}

\cmsNoteHeader{HIN-19-014}
\title{Observation of forward neutron multiplicity dependence of dimuon acoplanarity in ultraperipheral \texorpdfstring{\PbPb}{PbPb} collisions at \texorpdfstring{$\sqrtsNN=5.02\TeV$}{sqrt(sNN) = 5.02 TeV}}

\date{\today}

\abstract{
The first measurement of the dependence of $\gamma\gamma \to \PGmp\PGmm$ production on the multiplicity of neutrons emitted very close to the beam direction in ultraperipheral heavy ion collisions is reported. Data for lead-lead interactions at $\sqrtsNN = 5.02\TeV$, with an integrated luminosity of approximately 1.5\nbinv, are collected using the CMS detector at the LHC. The azimuthal correlations between the two muons in the invariant mass region $8 < m_{\PGm\PGm} < 60\GeV$ are extracted for events including 0, 1, or at least 2 neutrons detected in the forward pseudorapidity range $\abs{\eta} > 8.3$. The back-to-back correlation structure from leading-order photon-photon scattering is found to be significantly broader for events with a larger number of emitted neutrons from each nucleus, corresponding to interactions with a smaller impact parameter. This observation provides a data-driven demonstration that the average transverse momentum of photons emitted from relativistic heavy ions has an impact parameter dependence. These results provide new constraints on models of photon-induced interactions in ultraperipheral collisions. They also provide a baseline to search for possible final-state effects on lepton pairs caused by traversing a quark-gluon plasma produced in hadronic heavy ion collisions.
}

\hypersetup{%
    pdfauthor={CMS Collaboration},%
    pdftitle={Observation of forward neutron multiplicity dependence of dimuon acoplanarity in ultraperipheral Pb + Pb collisions at 5.02 TeV},%
    pdfsubject={CMS},%
    pdfkeywords={CMS, heavy ion, ultraperipheral collisions}
    }

    \maketitle

    The Lorentz-boosted electromagnetic (EM) fields surrounding relativistic heavy ions with large charges can be treated as a flux of quasireal photons~\cite{PhysRev.45.729, vonWeizsacker:1934nji} with the flux intensity proportional to the square of the ion charge. Therefore, ions accelerated at colliders can interact when their impact parameter ($b$) is greater than twice the nuclear radius ($R_{\mathrm{A}}$), via photon-photon and photon-nucleus processes, the so-called ultraperipheral collisions (UPCs)~\cite{Bertulani:1987tz, BAUR2002359, Bertulani:2005ru, Baltz:2007kq, Klein:2020fmr}. Photon-photon interactions can be used to test quantum electrodynamics (QED) and to search for physics beyond the standard model~\cite{Baltz:2009jk, Baur:2007fv, Adams:2004rz, Adam:2019mby, Afanasiev:2009hy, Abbas:2013oua, Sirunyan:2018fhl, Aaboud:2017bwk, Aad:2019ock, Bruce:2018yzs}. Photon-nucleus interactions probe the gluon distribution at small Bjorken $x$ in the nucleon or nucleus~\cite{Adler:2002sc, Adamczyk:2017vfu, Afanasiev:2009hy, Abelev:2012ba, Abbas:2013oua, TheALICE:2014dwa, Acharya:2019vlb, Khachatryan:2016qhq, Sirunyan:2018sav, Sirunyan:2019nog}.

    The momentum of emitted quasireal photons is predominantly along the beam direction and the transverse momentum (\pt) is
    small, typically less than 30\MeV~\cite{Bertulani:2005ru, Baltz:2007kq}. Therefore, the lepton pairs produced from leading-order 
    photon-photon scattering ($\gamma\gamma \to \ell^{+}\ell^{-}$) possess small pair \pt and are nearly back-to-back in the
    azimuthal angle ($\phi$). Recently, photon-photon~\cite{Adam:2018tdm, Aaboud:2018eph} and photon-nucleus~\cite{Adam:2015gba, STAR:2019yox} 
    processes have been observed at very low \pt in hadronic ($b<2R_{\mathrm{A}}$) heavy ion collisions. 
    Interestingly, a broadening of lepton pair azimuthal angle correlations (or, equivalently, an increase of lepton pair \pt)
    is observed in hadronic collisions compared to that from UPCs~\cite{Adam:2018tdm, Aaboud:2018eph}. In hadronic events, a deconfined state of partonic matter, known as the quark-gluon plasma (QGP), can be formed. Therefore, final-state EM modifications of lepton pairs inside a QGP medium have been proposed as possible interpretations of the broadening effect~\cite{Adam:2018tdm,Aaboud:2018eph,Klein:2018fmp}. The initial \pt of the lepton pairs depends on the overlap integral of the photon fluxes produced by the two nuclei, and as a result, the average pair \pt ($\MeanPt$) could depend on the $b$ between the two colliding ions. Although models of the flux of photons integrated over a given $b$ range have large uncertainties~\cite{Klein:2020jom,Zha:2018tlq,Klein:2020fmr}, a QED calculation~\cite{Zha:2018tlq} predicts larger $\MeanPt$ for smaller $b$ values. Such a larger $\MeanPt$ in the initial state would broaden the pair angular correlation, which could explain the effects observed in more central hadronic collisions.
    
    To disentangle possible contributions from initial- and final-state effects to the modifications observed in hadronic heavy ion collisions, 
    an experimental handle on the $b$ dependence of lepton pair production in UPCs is essential. The photon-photon interactions 
    can occur in conjunction with the excitation of one or both of the ions via photon absorption into giant dipole resonances or higher excited states~\cite{Bertulani:2005ru, Baltz:2007kq, Klein:2020fmr, Baltz:2009jk, Berman:1975tt, Broz:2019kpl}.
    The giant dipole resonances typically decay by emitting a single neutron, while higher excited states may emit two or more neutrons. These forward neutrons 
    have very low relative momentum with respect to their parent ions, and therefore approximately retain the beam rapidity. 
    The contribution of higher excitations becomes larger as $b$ gets smaller~\cite{Bertulani:2005ru, Baltz:2007kq, Klein:2020fmr, Baltz:2009jk}. 
    Therefore, the number of emitted neutrons detected in the forward region can be used to classify UPC events into different $b$ ranges. 

    This Letter reports the first measurement of the forward neutron multiplicity dependence of $\GGMuMu$ 
    production in the muon pair invariant mass region $8 < m_{\PGm\PGm} < 60\GeV$ in lead-lead (\PbPb) UPCs at a nucleon-nucleon 
    center-of-mass energy $\sqrtsNN = 5.02\TeV$, using data collected with the CMS detector during the 2018 LHC run. 
    The \PbPb sample that includes information about forward neutrons corresponds to an integrated luminosity of 
    approximately 1.5\nbinv. Azimuthal correlations of muon pairs, quantified by the acoplanarity, $\alpha = 1 - \abs{\phi^+ - \phi^-}/\pi$, 
    are presented for several different classes of neutron multiplicity detected in the forward pseudorapidity range $\abs{\eta} > 8.3$. Here, 
    $\phi^{\pm}$ represent the azimuthal angle of each muon in the lab frame. A larger average $\alpha$ for lepton pairs from leading-order $\TwoGamma$ scatterings corresponds to fewer back-to-back azimuthal correlations, and thus larger initial \pt of the interacting photons. The muon azimuthal angle is used instead of \pt because of its superior experimental resolution.
    The average invariant mass of muon pairs in various neutron multiplicity classes is also presented as a probe of the initial photon 
    energy and its $b$ dependence. Tabulated results are provided in the HEPData record for this analysis~\cite{hepdata}.
    
    The central feature of the CMS apparatus is a superconducting solenoid of 6\unit{m} internal diameter, providing a magnetic field of 3.8\unit{T}. 
    Within the solenoid volume, there are four subdetectors, including a silicon pixel and strip tracker detector, a lead-tungstate crystal
    electromagnetic calorimeter, and a brass and scintillator hadron calorimeter, each composed of a barrel and two end cap sections. Muons are
    detected in the range $\abs{\eta} < 2.4$ in gas-ionization detectors embedded in the steel flux-return yoke outside the 
    solenoid, with detection planes made using three technologies: drift tubes, cathode strip chambers, and resistive-plate chambers. 
    Matching muons to tracks measured in the silicon tracker leads to a relative \pt resolution around 1\%~\cite{Chatrchyan:2012xi} 
    and an azimuthal angle resolution better than $7\times 10^{-4}$ rad for a typical muon in this analysis. 
    The CMS experiment has extensive forward calorimetry, including two steel and quartz-fiber Cherenkov hadron forward (HF) calorimeters that 
    cover the range of $2.9 < \abs{\eta} < 5.2$, which are used to reject hadronic \PbPb collision events. Two zero degree calorimeters (ZDC)~\cite{Suranyi:2021ssd}, made of quartz fibers and plates embedded in tungsten absorbers, are used to detect neutrons from nuclear dissociation events in the range $\abs{\eta} > 8.3$. A detailed description of the CMS detector, together with a definition of the coordinate system used and the relevant kinematic variables, can be found in Ref.~\cite{Chatrchyan:2008aa}.

    Events used in this study were selected online using a hardware-based trigger system that requires at least one muon candidate coincident with a \PbPb bunch crossing~\cite{Khachatryan:2016bia}. On the trigger level, there is no explicit selection on the minimum muon \pt and events with an energy deposit above the noise threshold in both HF calorimeters are vetoed. For the off-line analysis, events have to pass a set of selection criteria designed to reject beam-related background processes (beam-gas collisions and beam scraping events) and hadronic collisions. Events are required to have a primary interaction vertex, formed by two or more tracks, within 20\unit{cm} from the CMS detector center along the beam axis. The cluster shapes in the pixel detector must be compatible with those expected from particles produced by a \PbPb collision~\cite{Chatrchyan:2011sx}.
    To suppress hadronic \PbPb collisions, the largest energy deposits in the HF calorimeters are required to be below 7.3 and 7.6\GeV in the positive and negative rapidity sides, respectively, where these noise thresholds are determined from empty bunch crossing events. In addition, events must contain exactly two muon candidates and no additional tracks in the range $\abs{\eta} < 2.4$. Selected events 
    are then classified by neutron multiplicity, which is determined by the energies deposited in the ZDCs. For single neutrons, the relative energy resolution of the ZDCs is $\sim$22\%--26\%, while the detection efficiency is close to 100\% in simulated events~\cite{Suranyi:2021ssd}. Based on neutron 
    peaks observed in the total ZDC energy distribution (\suppMaterial{}), events are
    divided into three neutron multiplicity classes ($0\Pn$, $1\Pn$, and $X\Pn$ with $X \geq 2$) on each side.
    The corresponding purities of selected neutron multiplicity classes are estimated by a multi-Gaussian function fit to the energy distribution. The purities are nearly 100\% for the 
    $0\Pn$ and $X\Pn$ classes, but only $\sim$93\%--95\% for the $1\Pn$ class because of detector resolution effects. From the combinations of the number of neutrons in each ZDC separately, a total of six neutron multiplicity classes, labeled as $0\Pn 0\Pn$, $0\Pn 1\Pn$, $0\Pn X\Pn$, $1\Pn 1\Pn$, $1\Pn X\Pn$,
    and $X\Pn X\Pn$, are used in this study. The $0\Pn 0\Pn$ class corresponds to no Coulomb breakup of either nucleus and the $1\Pn X\Pn$ class corresponds to one neutron emitted from one nucleus and at least two neutrons emitted from the other nucleus.

    Muons are selected in the kinematic range of $\pt^{\PGm} > 3.5\GeV$ and $\abs{\eta^{\PGm}} < 2.4$. They are reconstructed using the combined information of the tracker and muon detectors (so-called ``soft muons" defined in
    Ref.~\cite{Chatrchyan:2012xi}). 
    The opposite-sign distribution (signal and background) is reconstructed by combining 
    $\PGmp$ and $\PGmm$ candidates, while the combinatorial background is estimated using events containing same-sign muons. One of the muon candidates in the opposite- or same-sign pair is required to match a trigger muon. 
    The studied dimuon kinematic range is $8 < m_{\PGm\PGm} < 60\GeV$ and $\abs{y^{\PGm\PGm}} < 2.4$ to ensure high efficiency and also to
    suppress the contribution from photoproduced resonances (charmonia and \PZ bosons).

    The detector reconstruction efficiency is estimated using a dedicated $\GGMuMu$ Monte Carlo simulation 
    sample produced by the \textsc{STARlight} (v3.0) event generator~\cite{Klein:2016yzr} without restriction on the Coulomb breakup of either nucleus. 
    Only $\ell^{+}\ell^{-}$ pairs from the leading-order $\TwoGamma$ scattering are generated, and the calculation is performed
    by integrating over the entire $b$ space for UPC events. No differential $b$ dependence
    of the initial photon \pt is considered in \textsc{STARlight}. The CMS detector response is simulated further using \GEANTfour with these \textsc{STARlight} generated events~\cite{Agostinelli:2002hh}. The muon trigger ($\trigEff^{\PGm}$) and reconstruction ($\recoEff^{\PGm}$) efficiencies are estimated as functions of $\pt^{\PGm}$, $\eta^{\PGm}$, and $\phi^{\PGm}$. To correct for detector inefficiencies, each muon pair event is scaled by $(\trigEff\recoEff)^{-1}$, where $\trigEff = 1-(1-\trigEff^{\PGmp})(1-\trigEff^{\PGmm})$ and $\recoEff = \recoEff^{\PGmp}\recoEff^{\PGmm}$. The reconstruction and trigger efficiencies rapidly reach a plateau as functions of \pt with values of $\sim$95\%--99\% above $\pt^{\PGm} \approx 6\GeV$ for $\abs{\eta^{\PGm}} < 1.2$ and above $\pt^{\PGm} \approx 4\GeV$ for $1.2 < \abs{\eta^{\PGm}} < 2.4$. Systematic uncertainties associated with the efficiency corrections are negligible since they largely cancel out in the final observables, which are normalized by the total yield.

    The cross section of single electromagnetic dissociation (EMD)~\cite{Pshenichnov:2001qd, Pshenichnov:2011zz} of Pb nuclei in \PbPb collisions was measured to be $187.4\pm 0.2\stat ^{+13.2}_{-11.2}\syst\unit{b}$ at $\sqrtsNN = 2.76\TeV$~\cite{ALICE:2012aa}. It is expected to be even larger at $\sqrtsNN = 5.02\TeV$ given the stronger EM fields.
    Because of the large single-EMD cross section, a single measured $\GGMuMu$ event may contain concurrent EMD \PbPb events in the same bunch crossing. These concurrent events can emit neutrons and migrate the neutron multiplicity of a single 
    $\GGMuMu$ interaction to higher values. This EMD pileup effect is quantified by
    measuring the ZDC energy distributions from ``zero-bias" triggered events that require only the presence of both beams in the same 
    bunch crossing. No valid collision vertex or track is allowed to be present in the event. The same HF veto thresholds as for the $\GGMuMu$ events are applied. The neutron multiplicity classes in these selected zero-bias events are used to estimate the probability of a $\GGMuMu$ event being assigned an incorrect neutron multiplicity because of pileup effects. By inverting a matrix of these migration probabilities, the true observable distributions are extracted from the measured data. In this study, about 11\% of measured $\GGMuMu$ events have neutron multiplicity migration caused by EMD pileup. 

    \begin{figure*}[htbp]
        \centering
        \includegraphics[keepaspectratio,width=1\textwidth]{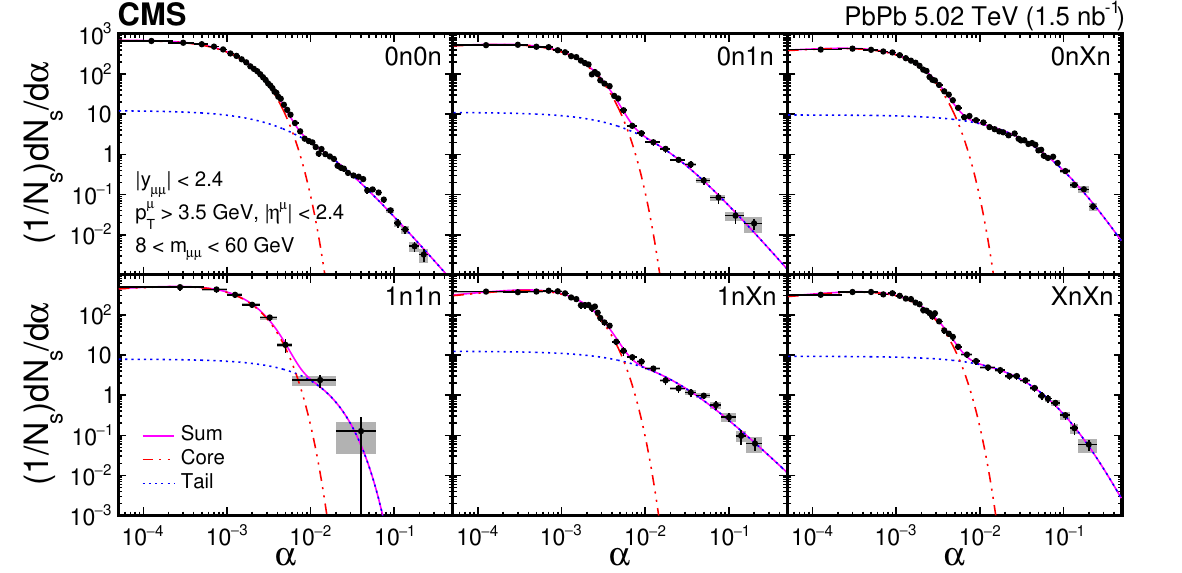}
        \caption{Neutron multiplicity dependence of acoplanarity distributions from $\GGMuMu$ for $\pt^{\PGm} > 3.5\GeV$, $\abs{\eta^{\PGm}} < 2.4$, $\abs{y^{\PGm\PGm}} < 2.4$, and $8 < m_{\PGm\PGm} < 60\GeV$ in ultraperipheral \PbPb collisions at $\sqrtsNN = 5.02\TeV$. 
        The $\alpha$ distributions are normalized to unit integral over their measured range. The dot-dot-dashed and dotted lines indicate the core and tail contributions, respectively, found using a fit to Eq.~(\ref{eq:LOHOfun}). The vertical lines on data
        points depict the statistical uncertainties, while the systematic uncertainties and horizontal bin widths are shown as gray boxes.}
        \label{fig:alphaSpec}
    \end{figure*}

    Figure~\ref{fig:alphaSpec} shows the corrected $\alpha$ distributions of $\PGmp\PGmm$ pairs in \PbPb collisions within the kinematic range ($\pt^{\PGm} > 3.5\GeV$, $\abs{\eta^{\PGm}} < 2.4$, and $\abs{y^{\PGm\PGm}} < 2.4$) for different neutron multiplicity classes. The $\alpha$ distributions 
    are normalized to unit integral over their measured range [$(1/N_{\mathrm{s}})\rd{N_{\mathrm{s}}}/\rd{\alpha}$, where $N_{\mathrm{s}}$ represents the signal yield]. Each $\alpha$ spectrum is characterized by a narrow core close to zero and a long tail. The core component mostly originates from the 
    leading-order $\TwoGamma$ scattering, while in the tail component, higher-order $\TwoGamma$ processes dominate. These higher-order
    processes include, \eg, extra photon radiation from the produced lepton(s), multiple-photon interactions, or scattering of (one or both) photons emitted from one of the protons inside the nucleus~\cite{Klein:2018fmp, Baur:2007fv}.
    The tail contribution in the $X\Pn X\Pn$ class is larger than that in the $0\Pn 0\Pn$ class. 
    This is consistent with the expectation of larger contributions of higher-order $\TwoGamma$ processes in UPC events that have smaller 
    $b$ and produce more neutrons in the forward region. 

    To investigate a possible $b$ dependence of the initial photon \pt, the core contribution to the $\alpha$ distribution 
    is decoupled from the tail contribution using a two-component empirical fit function
    (where $c_{i}$ and $t_{i}$ are the fit parameters),
    as shown in Fig.~\ref{fig:alphaSpec},
    \begin{linenomath}
        \begin{equation}
            \begin{aligned}
                \text{core}&:c_1  \re^{-\alpha/c_2 + c_3 \alpha^{0.25}}, \\
                \text{tail}&:t_{1} [1 + (t_2/t_3)  \alpha]^{-t_3},
                \label{eq:LOHOfun}
            \end{aligned}
        \end{equation}
    \end{linenomath}

    \noindent  except for the case of $1\Pn 1\Pn$, where a simple exponential function is used for the tail component, given the limited number of events. 
    The core component is largely modeled by an exponential function with a correction term ($c_3$) to account for the small depletion in the very small 
    $\alpha$ (\eg, $<5\times10^{-4}$) region, which tends to become more evident as the neutron multiplicity increases.
    This core functional form is validated by the \textsc{STARlight} event generator and 
    leading-order QED calculations, resulting in a $<$0.3\% discrepancy on the average acoplanarity from the fit and theoretical predictions. A binned $\chi^2$ goodness-of-fit minimization is performed using the integral of the function across each bin to account for the finite binning effect of the histogram. The average acoplanarity of $\PGmp\PGmm$ pairs from the core component ($\MeanAlphaCore$) is then calculated using the fit function.
    
    The measured $\alpha$ distribution and $\MeanAlphaCore$ of $\PGmp\PGmm$ pairs have several sources of systematic uncertainty arising from the contamination of hadronic collisions, EMD pileup correction, neutron multiplicity classification, and fit procedure. The uncertainty of the hadronic contamination is estimated by removing the requirement that selected events only contain two muons and is found to be $<$1.1\%. To estimate the systematic uncertainty associated with the HF noise threshold, the threshold to define the hadronic contamination is tightened to 5\GeV for both UPCs and zero-bias triggered events. The difference from the nominal result is quoted as the systematic uncertainty and contributes $<$2.7\%. The uncertainty arising from impure $1\Pn$ class selection ($<$0.7\%) is estimated by subtracting the contributions of $2\Pn$ events selected with tight energy requirements, according to the $2\Pn$ contamination probability. 
    The systematic uncertainty associated with contamination of photoproduced \PgU mesons ($\sim$0.6\%) is estimated
    by comparing $\alpha$ distributions from \textsc{STARlight} between pure $\GGMuMu$ and 
    $\GGMuMu$ mixed with photoproduced coherent \PgUa, with the relative yield ratio of \PgUa over 
    $\GGMuMu$ estimated by fitting the invariant mass distribution. The systematic uncertainty in $\MeanAlphaCore$ associated with the binned $\chi^2$ fit procedure is estimated by varying the bin width of $\alpha$ distributions, and is found to be less than 4\%. The total systematic uncertainties are derived 
    from a quadratic sum of all systematic sources and are found to be at most 5.1\% in $\MeanAlphaCore$. To measure $\MeanMass$, a second-order polynomial function is fit to the mass spectrum (\suppMaterial{}), to interpolate the contribution of $\TwoGamma$ scattering to dimuon pair production over the $\PgU$ mass region. The systematic uncertainty related to this procedure is estimated by comparing the nominal result to the one obtained by a third-order polynomial function fit. Together with the aforementioned systematic sources, the total systematic uncertainty in $\MeanMass$ is below 1.8\%, across all neutron multiplicity classes.

    \begin{figure}[htbp]
        \centering
        \includegraphics[width=0.49\textwidth]{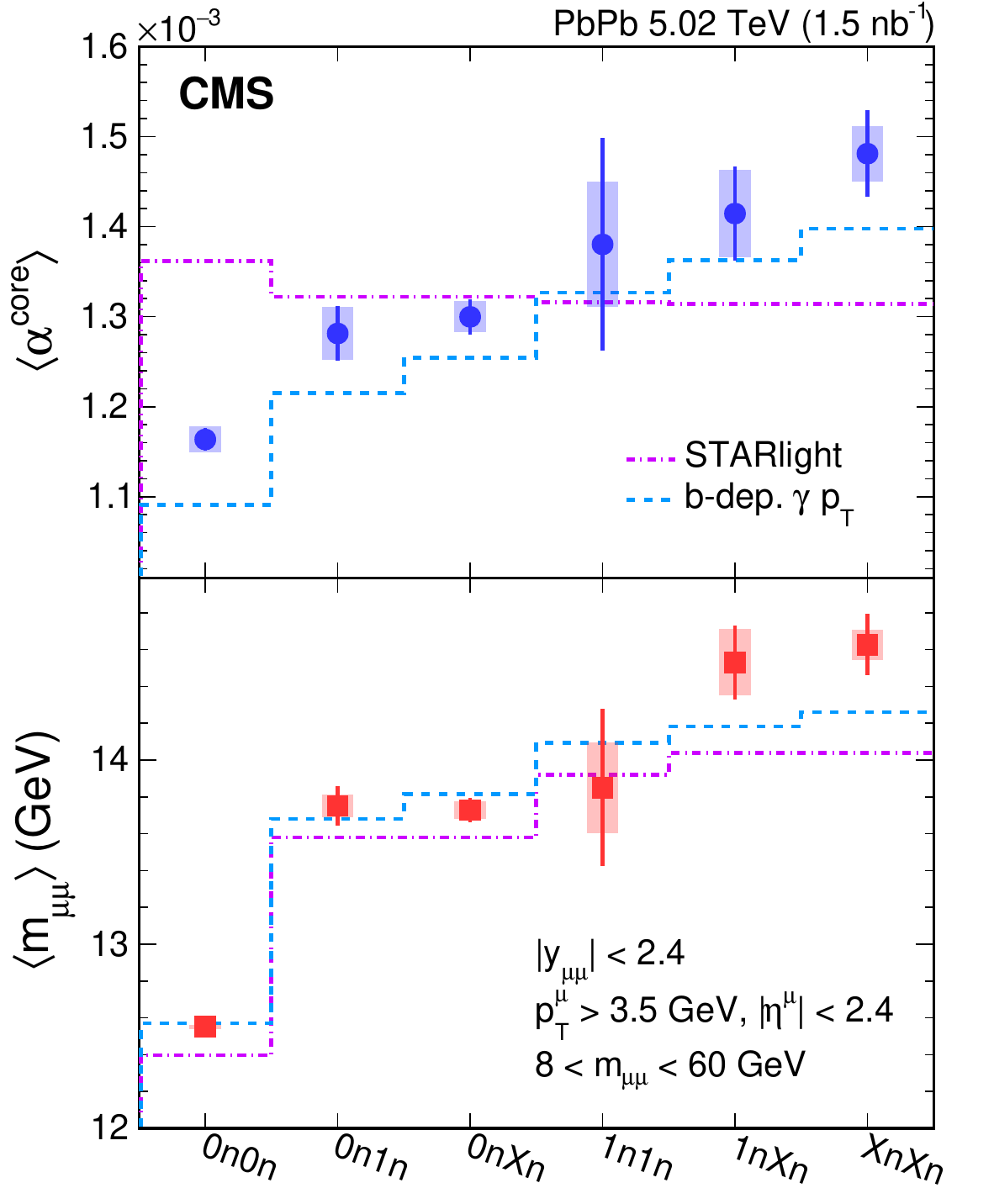}
        \caption{Neutron multiplicity dependence of (upper) $\MeanAlphaCore$ and (lower) $\MeanMass$ of $\PGmp\PGmm$ pairs in ultraperipheral \PbPb collisions at $\sqrtsNN = 5.02\TeV$. The vertical lines on data points depict the statistical uncertainties, while the systematic uncertainties of the data are shown as shaded areas. The dot-dashed line shows the \textsc{STARlight} prediction, and the dashed line corresponds to the leading-order QED calculation of Ref.~\cite{Brandenburg:2020ozx}.}
        \label{fig:meanAlphaAndMassvsNeuNum}
    \end{figure}

    The neutron multiplicity dependence of $\MeanAlphaCore$ for $\PGmp\PGmm$ pairs in ultraperipheral \PbPb collisions at $\sqrtsNN = 5.02\TeV$ is shown in Fig.~\ref{fig:meanAlphaAndMassvsNeuNum} (upper), in the mass region $8 < m_{\PGm\PGm} < 60\GeV$. A strong neutron multiplicity dependence of $\MeanAlphaCore$ is clearly observed, while the $\MeanAlphaCore$ predicted by \textsc{STARlight} is almost constant at a value of about $1.35\times 10^{-3}$, shown as the dot-dashed line in Fig.~\ref{fig:meanAlphaAndMassvsNeuNum} (upper). The $\MeanAlphaCore$ for inclusive UPCs is measured to be $[1227\pm 7\stat\pm 8\syst]\times 10^{-6}$, about 10\% lower than the \textsc{STARlight} prediction. In general, the $\MeanAlphaCore$ in data becomes larger as the emitted neutron multiplicity increases.
    A fit to the dependence of $\MeanAlphaCore$ on the neutron multiplicity with a constant value is rejected with 
    a $p$ value corresponding to 5.7 standard deviations.
    This observation demonstrates that initial photons producing 
    $\PGmp\PGmm$ pairs have a significant $b$ dependence of their \pt, which impacts the \pt and acoplanarity 
    of muon pairs in the final state. This initial-state contribution must be properly taken into account when exploring possible final-state 
    EM effects arising from a hot QGP medium formed in hadronic heavy ion collisions~\cite{Adam:2018tdm, Aaboud:2018eph}. 
    A recent leading-order QED calculation~\cite{Brandenburg:2020ozx}, incorporating a $b$ dependence of the initial photon \pt~\cite{Zha:2018tlq}, 
    has provided results for all the reported neutron multiplicity classes. The average $b$ values estimated in Ref.~\cite{Brandenburg:2020ozx} range 
    from about 112 to 22\unit{fm} for the $0\Pn 0\Pn$ to $X\Pn X\Pn$ neutron multiplicity classes, respectively. The model calculation
    can qualitatively describe the increasing trend of $\MeanAlphaCore$ data with the neutron multiplicity, 
    shown as the dashed line in Fig.~\ref{fig:meanAlphaAndMassvsNeuNum} (upper). However, the data are systematically higher than the 
    model calculation (plotted without uncertainties) by about 5\%, which may be related to the presence in data of soft photon radiation from the muons~\cite{Klein:2018fmp}.

\begin{figure*}[htbp]
    \centering
    \includegraphics[width=1\textwidth]{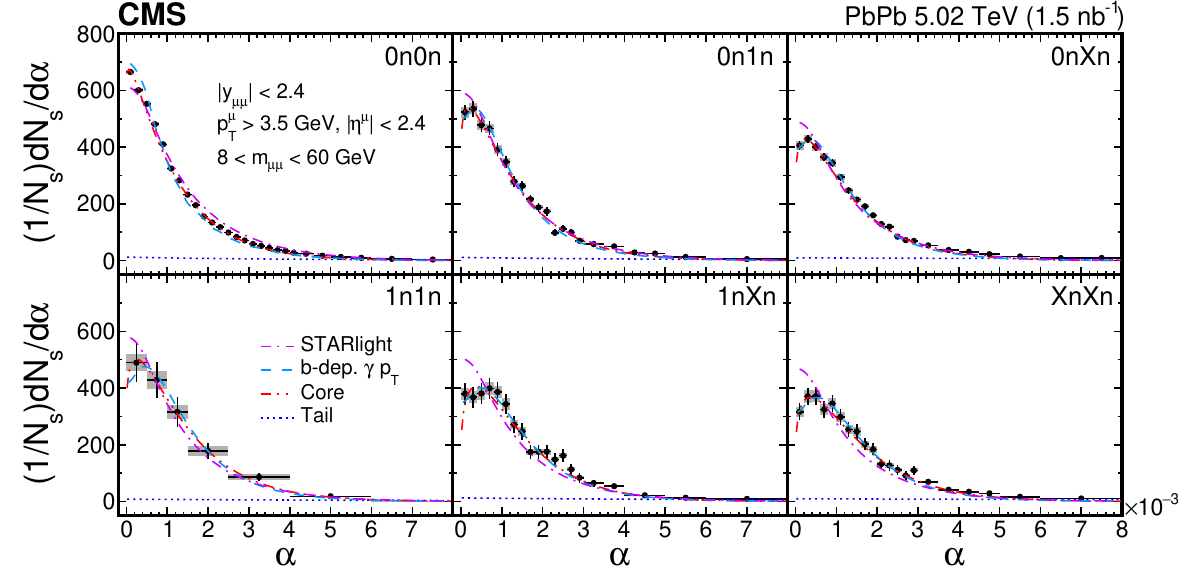}
    \caption{A direct comparison between data and model calculations for the $\alpha$ distributions with $\alpha < 0.008$. The dot-dashed line shows the \textsc{STARlight} prediction and the dashed line corresponds to the leading-order QED calculation of Ref.~\cite{Brandenburg:2020ozx}. The dot-dot-dashed and dotted lines indicate the core and tail contributions, respectively. The vertical lines and shaded areas represent the statistical and systematic uncertainties, respectively.}
    \label{fig:coreAlphaSpec}
\end{figure*}

    Figure~\ref{fig:coreAlphaSpec} shows a direct comparison between data and model calculations for the $\alpha$ distributions with $\alpha < 0.008$ in all the reported neutron multiplicity classes. The $\alpha$ distribution is clearly observed to broaden in the high neutron multiplicity class. The QED calculation, incorporating a $b$ dependence of the initial photon \pt, can describe the $\alpha$ distributions reasonably well while \textsc{STARlight} fails to describe the data.

    A rapidity dependence of the $\alpha$ distribution is also investigated for $0\Pn 1\Pn$, $0\Pn X\Pn$ and $1\Pn X\Pn$ classes 
    (\suppMaterial{}) for dimuon rapidity in the hemisphere containing larger (smaller) neutron multiplicity. In the $0\Pn X\Pn$ class, the tail contribution in the rapidity 
    hemisphere with $X\Pn$ is significantly larger than that in the rapidity hemisphere with zero neutrons, suggesting 
    contributions from different higher-order processes that correlate with the dimuon pair production. However, no rapidity dependence is observed 
    for the $\MeanAlphaCore$ values extracted from the fits using Eq.~(\ref{eq:LOHOfun}).
    This is consistent with the expectation that the $\MeanAlphaCore$ is dominated by leading-order 
    $\GGMuMu$ scatterings and illustrates that the employed core functional form is robust.
        
    In Fig.~\ref{fig:meanAlphaAndMassvsNeuNum} (lower), the average invariant mass $\MeanMass$ of all muon pairs passing the selection criteria, is shown as a function of the neutron multiplicity. A clear neutron multiplicity dependence 
    of $\MeanMass$ is observed, with the $\MeanMass$ value measured in $X\Pn X\Pn$ events being larger than that 
    in $0\Pn 0\Pn$ events with a significance exceeding 5 standard deviations. This trend of $\MeanMass$ can be qualitatively described by both model calculations. As the muon pair invariant mass is largely determined by the initial photon energy, this observation suggests that the energy of the photons involved in UPCs is, on average, larger in collisions with smaller $b$, 
    a conclusion similar to that previously drawn for the initial photon \pt.

    In summary, the first measurements of $\GGMuMu$ production as a function of forward 
    neutron multiplicity in ultraperipheral lead-lead collisions at a nucleon-nucleon center-of-mass energy of 5.02\TeV are reported. 
    A significant broadening of back-to-back azimuthal correlations is seen, with respect to the leading-order $\GGMuMu$ process, for increasing multiplicities of emitted forward neutrons. This observed trend is qualitatively reproduced by a leading-order quantum electrodynamics calculation, demonstrating the importance of an impact-parameter-dependent photon \pt . A similar trend of increasing average invariant
    mass of muon pairs with neutron multiplicity is also observed. 
    These measurements provide the first experimental demonstration that the initial energy and transverse momentum of photons exchanged in ultraperipheral heavy ion collisions depend on the impact parameter of the interaction. These results call for theoretical efforts to improve the 
    precision in modeling photon-induced interactions. Future searches for electromagnetic interactions of leptons inside the quark-gluon plasma created in heavy ion collisions should incorporate a baseline where the initial broadening effects presented in this Letter are properly taken into account.

    \begin{acknowledgments}
        We congratulate our colleagues in the CERN accelerator departments for the excellent performance of the LHC and thank the technical and administrative staffs at CERN and at other CMS institutes for their contributions to the success of the CMS effort. In addition, we gratefully acknowledge the computing centers and personnel of the Worldwide LHC Computing Grid for delivering so effectively the computing infrastructure essential to our analyses. Finally, we acknowledge the enduring support for the construction and operation of the LHC and the CMS detector provided by the following funding agencies: BMBWF and FWF (Austria); FNRS and FWO (Belgium); CNPq, CAPES, FAPERJ, FAPERGS, and FAPESP (Brazil); MES (Bulgaria); CERN; CAS, MoST, and NSFC (China); COLCIENCIAS (Colombia); MSES and CSF (Croatia); RIF (Cyprus); SENESCYT (Ecuador); MoER, ERC PUT and ERDF (Estonia); Academy of Finland, MEC, and HIP (Finland); CEA and CNRS/IN2P3 (France); BMBF, DFG, and HGF (Germany); GSRT (Greece); NKFIA (Hungary); DAE and DST (India); IPM (Iran); SFI (Ireland); INFN (Italy); MSIP and NRF (Republic of Korea); MES (Latvia); LAS (Lithuania); MOE and UM (Malaysia); BUAP, CINVESTAV, CONACYT, LNS, SEP, and UASLP-FAI (Mexico); MOS (Montenegro); MBIE (New Zealand); PAEC (Pakistan); MSHE and NSC (Poland); FCT (Portugal); JINR (Dubna); MON, RosAtom, RAS, RFBR, and NRC KI (Russia); MESTD (Serbia); SEIDI, CPAN, PCTI, and FEDER (Spain); MOSTR (Sri Lanka); Swiss Funding Agencies (Switzerland); MST (Taipei); ThEPCenter, IPST, STAR, and NSTDA (Thailand); TUBITAK and TAEK (Turkey); NASU (Ukraine); STFC (United Kingdom); DOE and NSF (U.S.).    \end{acknowledgments}

    \bibliography{auto_generated}

    \numberwithin{figure}{section}
    \ifthenelse{\boolean{cms@external}}{\nocite{Zyla:2020zbs}}{
    \clearpage\appendix
    \section{ZDC energy distributions and rapidity dependence of acoplanarity distributions and subtraction of \texorpdfstring{\PgU}{PgU} mesons \label{app:suppMat}}
    Figure~\ref{fig:zdc} (left) shows the correlation between energy distributions of the ZDC detectors, located on the positive (Plus) and negative (Minus) directions with respect to the CMS interaction point, for events selected in the analysis. Figure~\ref{fig:zdc} (right) shows the measured Minus ZDC energy distribution together with a multi-Gaussian function fit.

\begin{figure*}[htbp]
\centering
\includegraphics[width=1\textwidth]{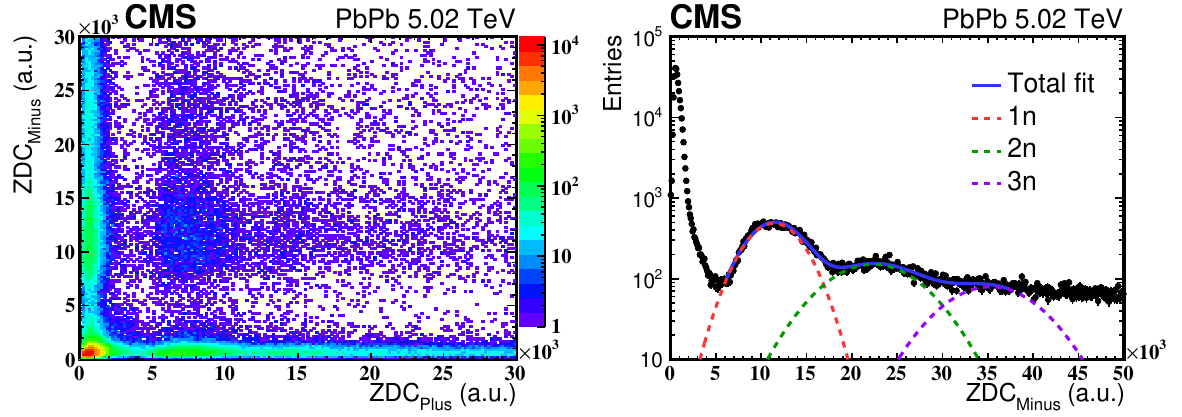}
    \caption{The left panel shows the correlation between energy distributions of the Minus and Plus ZDC detectors (one entry per event), while the right panel shows a multi-Gaussian function fit to the Minus ZDC energy distribution.}
    \label{fig:zdc}
\end{figure*}

For the measured neutron multiplicity class with asymmetric neutron numbers, the dimuon rapidity is divided into two hemispheres using the plane defined by $y = 0$. In each rapidity hemisphere, the $\alpha$ distribution from $\GGMuMu$ is normalized by the total yields in this neutron multiplicity class ($(1/N_{\mathrm{s}})\rd{N_{\mathrm{s}}^{\mathrm{rap}}}/\rd{\alpha}$, where the $N_{\mathrm{s}}$ represents the total yields and $N_{\mathrm{s}}^{\mathrm{rap}}$ represents the yields in each rapidity hemisphere), as shown in Fig.~\ref{fig:rapAlpha}.

\begin{figure*}[htbp]
    \centering
    \includegraphics[width=1\textwidth]{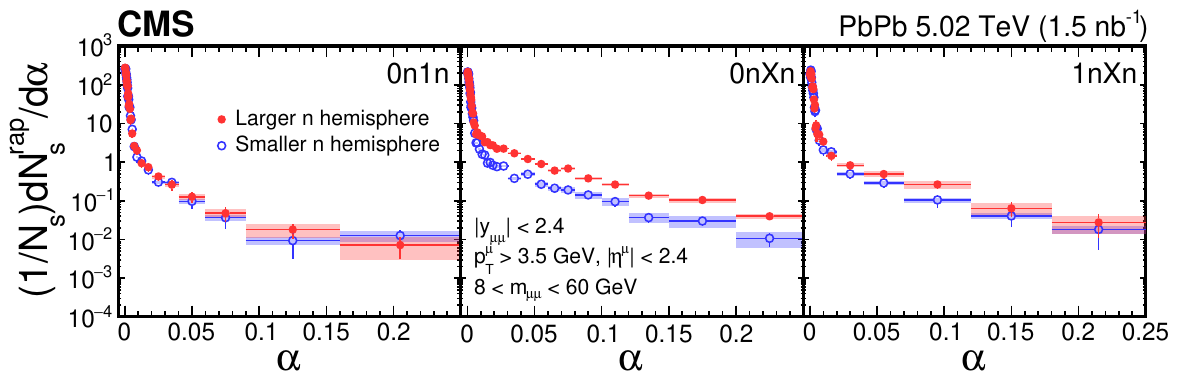}
    \caption{Acoplanarity distributions of $\GGMuMu$ events for three different neutron multiplicity classes with asymmetric neutron numbers. The solid red (open blue) symbols correspond to events where the dimuon rapidity is in the hemisphere containing larger (smaller) neutron multiplicity. The vertical lines on data points depict the statistical uncertainties while the systematic uncertainties are shown as shaded areas.}
    \label{fig:rapAlpha}
\end{figure*}

The yields of muon pairs from $\TwoGamma$ scattering in the $\PgU$ mass region ($9 < m_{\PGm\PGm} < 11\GeV$) are extracted by a binned $\chi^2$ fit to the invariant mass spectrum, as shown in Fig.~\ref{fig:massSpec}. Each $\PgU$ state is modeled by a Gaussian function. All the parameters of the $\PgUa$ fit are left free. For the $\PgUb$ and $\PgUc$ states, the yields are allowed to vary while the mean and width are fixed to values found by multiplying those for $\PgUa$ by the ratio of the published masses of the states~\cite{Zyla:2020zbs}. The contribution of $\TwoGamma$ scattering to dimuon pair production in the $\PgU$ mass region is extracted using a second order polynomial function.

\begin{figure*}[htbp]
    \centering
    \includegraphics[width=0.6\textwidth]{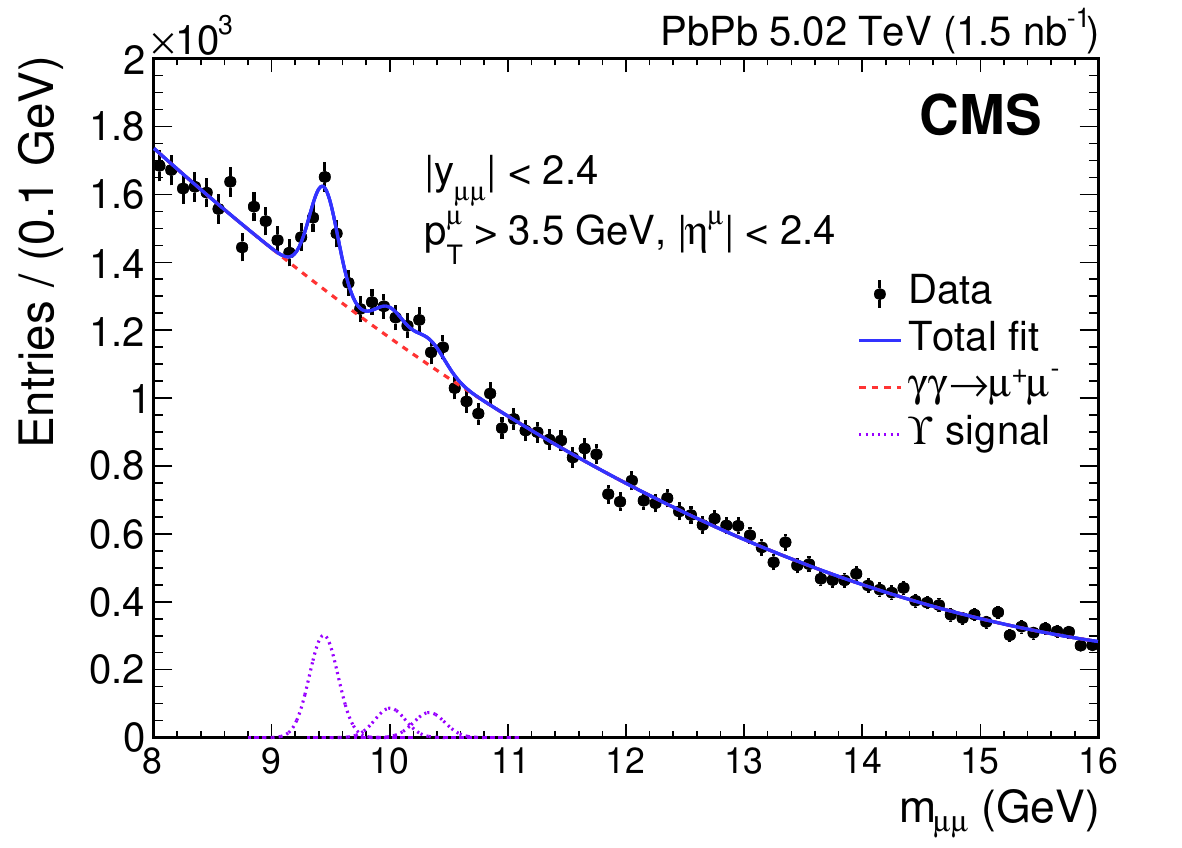}
    \caption{The efficiency corrected invariant mass distribution of muon pairs in inclusive ultraperipheral \PbPb collisions, for the kinematic range $\pt^{\PGm} > 3.5\GeV$, $\abs{\eta^{\PGm}} < 2.4$, and $\abs{y^{\PGm\PGm}} < 2.4$. The result of the fit to the data is shown as solid blue line. The yields of muon pairs from $\TwoGamma$ scattering in the $\PgU$ mass region are shown as dashed red line. The separate yields for each $\PgU$ state are shown as dotted violet lines.}
    \label{fig:massSpec}
\end{figure*}

    }

\cleardoublepage \section{The CMS Collaboration \label{app:collab}}\begin{sloppypar}\hyphenpenalty=5000\widowpenalty=500\clubpenalty=5000\vskip\cmsinstskip
\textbf{Yerevan Physics Institute, Yerevan, Armenia}\\*[0pt]
A.M.~Sirunyan$^{\textrm{\dag}}$, A.~Tumasyan
\vskip\cmsinstskip
\textbf{Institut f\"{u}r Hochenergiephysik, Wien, Austria}\\*[0pt]
W.~Adam, T.~Bergauer, M.~Dragicevic, J.~Er\"{o}, A.~Escalante~Del~Valle, R.~Fr\"{u}hwirth\cmsAuthorMark{1}, M.~Jeitler\cmsAuthorMark{1}, N.~Krammer, L.~Lechner, D.~Liko, T.~Madlener, I.~Mikulec, F.M.~Pitters, N.~Rad, J.~Schieck\cmsAuthorMark{1}, R.~Sch\"{o}fbeck, M.~Spanring, S.~Templ, W.~Waltenberger, C.-E.~Wulz\cmsAuthorMark{1}, M.~Zarucki
% ...
\vskip\cmsinstskip
\dag: Deceased\\
1:  Also at Vienna University of Technology, Vienna, Austria\\
% ...
\end{sloppypar}
%%% END EDITABLE REGION %%%
% skeleton_end
\end{document}